\documentclass{article} % For LaTeX2e
\usepackage[final]{colm2026_conference}

\usepackage{microtype}
\usepackage{hyperref}
\usepackage{url}
\usepackage{booktabs}
\usepackage{graphicx}

\usepackage{lineno}

\definecolor{darkblue}{rgb}{0, 0, 0.5}
\hypersetup{colorlinks=true, citecolor=darkblue, linkcolor=darkblue, urlcolor=darkblue}

\title{Grounded in Consensus, In Step With Emerging Science: A Consensus-Anchored Multi-Corpus Clinical Chatbot for Long COVID}

\author{Yining Wu \\
School of Information\\
University of Texas at Austin\\
Austin, TX, USA \\
\texttt{\{yining.wu\}@utexas.edu} \\
\And
Philip DiGiacomo \\
Department of Computer Science \\
University of Texas at Austin \\
Austin, TX, USA \\
\texttt{\{pdigiacomo\}@utexas.edu} \\
\AND
Ying Ding\\
School of Information, Dell Medical School \\
University of Texas at Austin
Austin, TX, USA\\
\texttt{\{ying.ding\}@ischool.utexas.edu}
\AND
William Brode\\
Dell Medical School \\
University of Texas at Austin
Austin, TX, USA\\
\texttt{\{\href{mailto:William.Brode@austin.utexas.edu}{William.Brode}\}@austin.utexas.edu}
}

\begin{document}

\ifcolmsubmission

\fi

\maketitle

\begin{abstract}
Long COVID (LC) poses a challenge for clinical decision support because relevant evidence is distributed across sources with different update cycles, evidentiary roles, and levels of clinical maturity. We present a clinician-facing chatbot that organizes four sources within a retrieval-augmented workflow: expert-curated consensus guidance, current PubMed literature, registered interventional trials, and evidence from living systematic reviews. Consensus guidance is always included to frame responses, while the remaining sources are retrieved in parallel when selected by the user. In an exploratory automated evaluation on 50 clinician-facing questions, our chatbot showed comparable mean ratings to OpenEvidence, with numerically higher scores and lower score variability in an LLM-judged comparison.

\end{abstract}

\section{Introduction}

Clinical decision-support systems for emerging diseases must do more than retrieve relevant information. Long COVID (LC) is a useful stress test because it combines high evidence volume with limited clinical certainty: heterogeneous multisystem symptoms, no reliable biomarker, no proven disease-modifying therapy, and rapidly changing literature. In deployment, such systems must determine which evidence sources should inform a response, how sources with different authority and update cycles should be combined, and how clinicians can inspect the provenance of resulting claims. LC clinical interpretation therefore draws on consensus guidance, emerging primary studies, registered trials, and continually updated evidence syntheses \cite{national2024long, zeraatkar2024interventions}.

Prior LC question-answering work found that broad literature retrieval can produce clinically misaligned outputs, while a curated corpus of clinical guidance and high-quality reviews produced more appropriate responses than PubMed-scale retrieval alone \cite{digiacomo2025demoguideragevidencedrivencorpus}. That result motivates a deployment problem: a static curated corpus can anchor responses, but cannot capture published findings, trial-status changes, or living-review updates. Nor should these sources simply be pooled, because consensus guidance, primary literature, trial registries, and systematic reviews serve distinct clinical roles.

We developed a clinician-facing chatbot that organizes these sources within a single retrieval-augmented workflow. Expert-curated consensus guidance is always included to frame responses, while PubMed literature, ClinicalTrials.gov records, and living systematic-review evidence are retrieved in parallel when selected by the user. This deployment case report describes the system and the design choices required to operationalize multi-source evidence synthesis in a clinical workflow. We focus on source-specific retrieval, consensus-anchored synthesis, provenance presentation, and reliability under real query load.

\section{Methods}

\subsection{System overview}

\begin{figure}[t]
    \centering
    \includegraphics[width=1\textwidth]{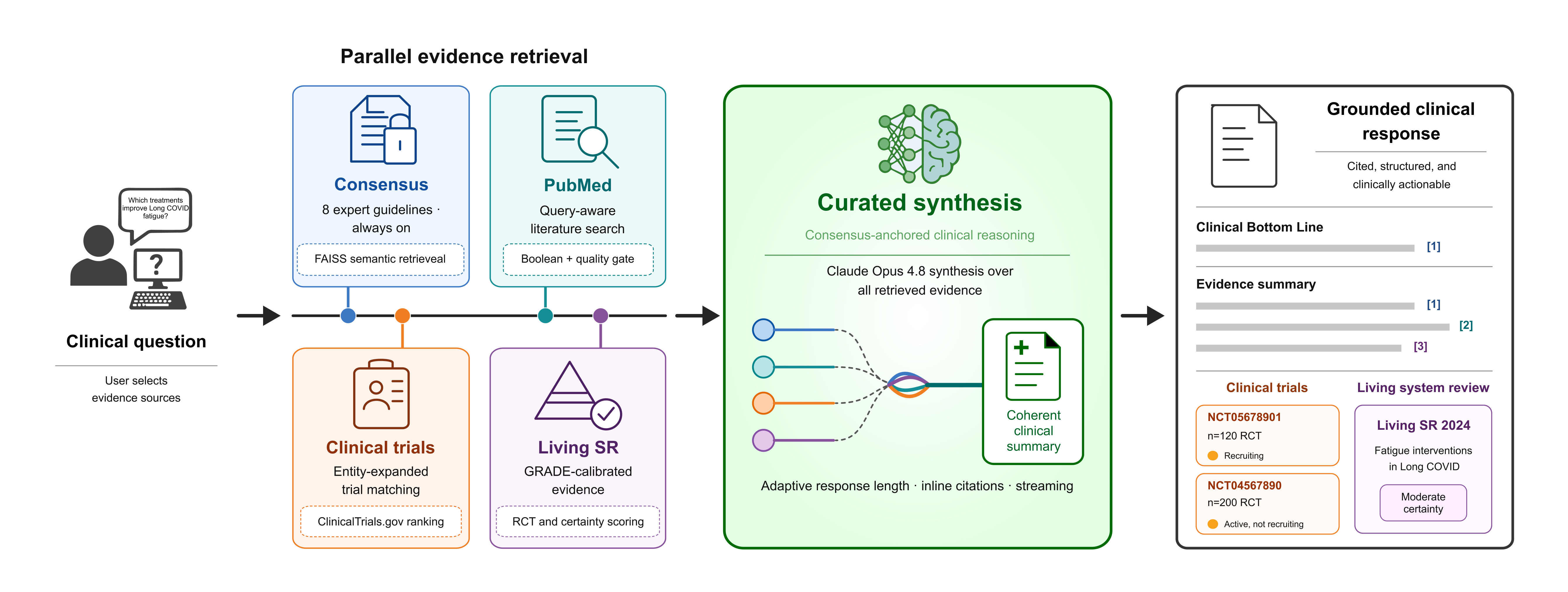}
    \caption{Four evidence sources (Consensus, PubMed, Clinical trials, and Living SR) are retrieved in parallel and integrated through a synthesis module into a clinical response.}
    \label{fig:pace}
\end{figure}

We developed a clinician-facing retrieval-augmented generation system that organizes four complementary evidence corpora into a single clinical workflow (Figure \ref{fig:pace}): expert consensus guidance (Consensus), PubMed literature (PubMed), ClinicalTrials.gov clinical trial records (Clinical trials), and living systematic review evidence (Living SR). Expert consensus guidance is always included as the foundational corpus, whereas PubMed literature, clinical trials, and Living SR evidence are user-selectable and enabled by default. Selected sources are retrieved in parallel and passed to a single Curated synthesis model, which produces a citation-traceable clinical response. The system is deployed as a clinician-facing web application with access intentionally limited to authorized users during the current deployment phase to meet security and operational requirements.

\subsection{Evidence corpora and retrieval}

The consensus corpus contains eight expert-curated LC guidance and consensus documents.
Documents were processed into semantic chunks and indexed using text-embedding-3-large; relevant passages were retrieved by vector similarity for each query.

Three additional corpora provide complementary evidence. The PubMed component retrieves current literature through NCBI E-utilities. An LLM-assisted query module identifies the clinical intent and relevant study characteristics, constructs a Boolean query with LC synonym constraints, and applies publication-type and recency filters before returning structured article findings.

The clinical-trials component indexes a curated registry corpus derived from ClinicalTrials.gov (currently 887 records). It uses query-derived clinical entities to rank relevant records and returns a shortlist with trial status, design characteristics, and available results summaries. This component is intended to characterize the LC trial landscape, including what interventions are being studied and the status of ongoing work.

The Living SR component indexes 74 randomized controlled trial publications included in living systematic reviews across eight intervention categories. Retrieved studies are linked to intervention--outcome-specific GRADE certainty assessments \citep{grade2004grading}, allowing the system to present trial-level findings and the certainty of the evidence.

Detailed corpus-construction procedures and source-specific retrieval configurations are provided in Appendix A.

\subsection{Synthesis, response processing, and reliability}

\begin{figure}[t]
    \centering
    \includegraphics[width=1\textwidth]{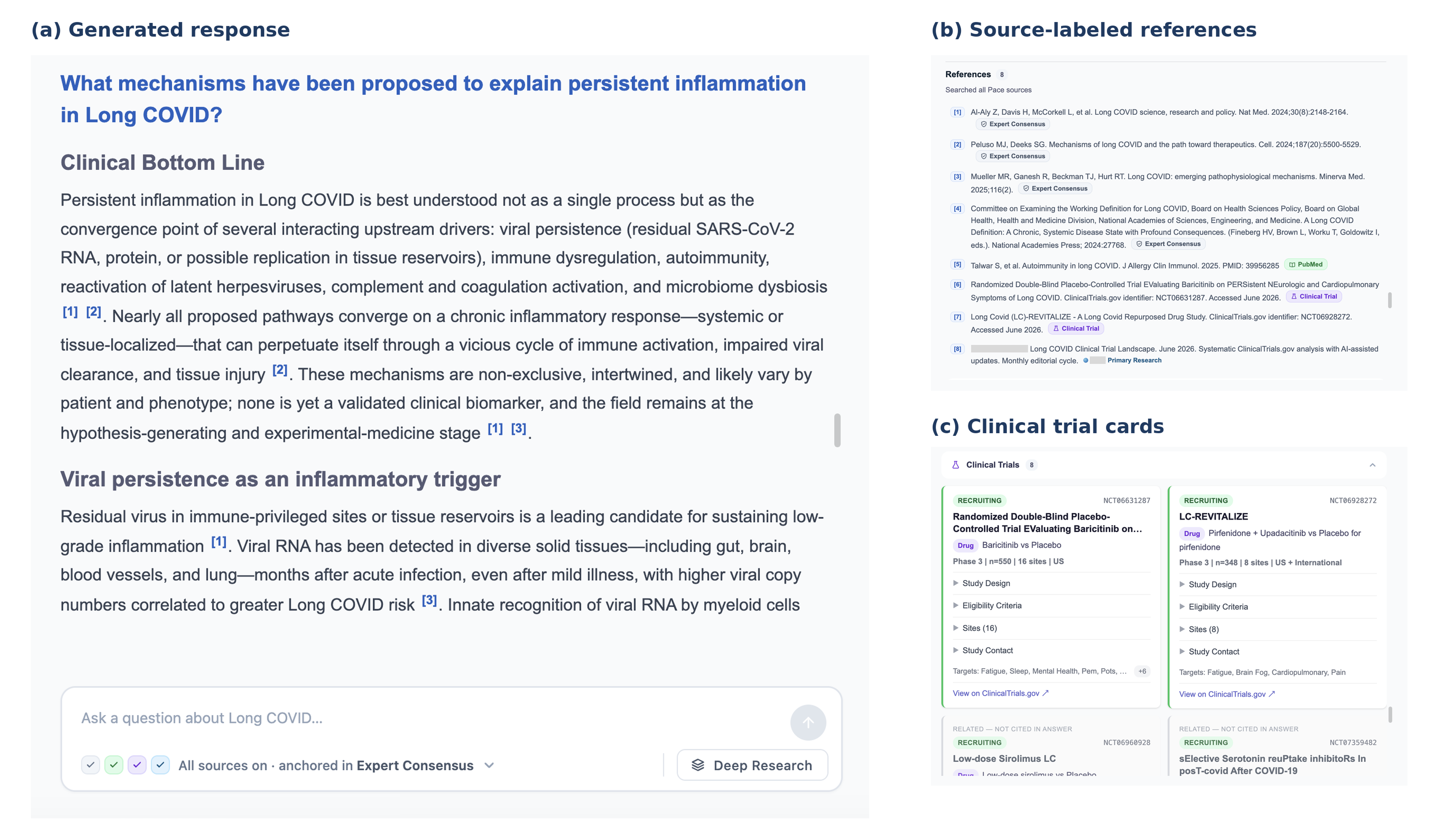}
    \caption{User-facing presentation of a curated synthesis response. (a) Generated responses begin with a section of Clinical Bottom Line and include inline citations. (b) A source-labeled reference list preserves provenance across evidence corpora. (c) Relevant ClinicalTrials.gov records are displayed as trial cards. }
    \label{fig:interface}
\end{figure}

Retrieved evidence is integrated by a single synthesis model (Claude Opus 4.8), which frames responses around the consensus corpus while incorporating findings from selected evidence streams where relevant. Responses begin with a Clinical Bottom Line and include inline source markers. An application-layer post-processing step converts these markers into numbered citations, de-duplicates overlapping PubMed and Living SR records, and renders a source-labeled reference list and evidence cards in the user interface (Figure \ref{fig:interface}).

Retrieval arms operate independently so that a timeout or failure in one source does not block the overall response; such failures are represented in the response metadata and interface. User-selected PubMed filters are enforced as hard constraints. When no records meet the selected criteria, the system reports an empty result rather than automatically relaxing the filters. The default workflow was designed for interactive clinical use, with a target latency below 45 seconds and a median latency of 29 seconds in internal testing.

\section{Evaluation}

We conducted an exploratory LLM-as-a-judge comparison of our chatbot and OpenEvidence on 50 LC-related questions selected by 6 domain experts. OpenEvidence was selected as a clinician-facing, evidence-based AI platform that provides cited point-of-care responses and has been evaluated in primary care settings \cite{hurt2025use}. GPT-4o independently scored each response on Factual Accuracy, Completeness and Thoroughness, and Clinical Soundness using evaluation dimensions adapted from prior work \cite{li2026clicare}. Scores were assigned on a 1--5 Likert scale, with higher values indicating better performance.

Mean ratings were numerically higher and score variability lower for our chatbot across all three dimensions (Table~\ref{eva}). In addition, no response from our chatbot received the lowest possible rating on any dimension, whereas three OpenEvidence responses received at least one lowest-rating score. Overall, the evaluation indicated comparable automated response quality, with less variable scores for our chatbot.

\begin{table}
\begin{center}
\begin{tabular}{lll}
\toprule
\multicolumn{1}{c}{\bf METRIC}  &\multicolumn{1}{c}{\bf Our chatbot} & \bf OpenEvidence\\
\midrule
Factual Accuracy&4.32 ± 0.68 &4.30 ± 1.05 \\
Completeness \& Thoroughness&4.22 ± 0.51 &4.18 ± 0.75 \\
Clinical Soundness
&4.38 ± 0.57 &4.34 ± 1.02 \\
\midrule
 Mean across dimensions
& 4.31 ± 0.55
&4.27 ± 0.91
\\
 Questions with at least one score of 1& 0 (0\%)&3 (6\%)\\ \bottomrule
\end{tabular}
\end{center}
\caption{GPT-4o evaluation of Long COVID responses from our chatbot and OpenEvidence. Values are mean ± SD across 50 questions. Higher scores indicate better performance.}\label{eva}
\end{table}

\section{Discussion}

Deploying a clinical support chatbot for LC requires balancing evidence breadth, uncertainty communication, response usability, and operational constraints. In this setting, the failure mode is not only missing evidence; it is also presenting early or low-certainty findings with more confidence than clinical consensus supports. Broader and more recent retrieval can improve coverage, but responses must remain interpretable within an interactive workflow; richer multi-stage processing can add context, but also increases latency and system complexity. These trade-offs shape whether a system functions as a usable clinical tool rather than a general-purpose search interface.

Our central design choice was to preserve, rather than pool, the clinical roles of evidence sources: consensus guidance frames interpretation, PubMed captures emerging findings, trial registries track ongoing investigation, and living reviews contextualize trial findings by certainty. The interface makes these distinctions inspectable through source selection, explicit empty-result and failure reporting, a cited Clinical Bottom Line, and source-specific evidence cards. Together, these choices position evidence orchestration, provenance, presentation, and latency as co-equal requirements for clinical AI in rapidly evolving evidence settings.

\section{Limitations and Future Work}
This report describes an LC chatbot deployment and exploratory automated evaluation on a limited set of clinician-facing questions. The LLM-judge evaluation did not assess patient outcomes, clinician behavior, real-time use, or the contribution of individual system components; it was not powered to detect small score differences. Results should therefore be interpreted as feasibility signals rather than evidence of superiority. We are conducting a human expert evaluation in which six Long COVID experts develop reference answers and review outputs from our chatbot, OpenEvidence, and frontier models. This review will assess clinical quality and help calibrate automated monitoring as models, prompts, and evidence evolve.

\section{Conclusion}

This deployment case report describes a clinician-facing LC chatbot that organizes foundational expert guidance, recent literature, registered trials, and living systematic-review evidence within a source-traceable workflow. Rather than treating retrieval as a single-corpus problem, the system preserves the complementary clinical roles of these evidence streams while balancing response depth, interaction speed, and provenance presentation. Our experience illustrates how evidence orchestration can support deployable clinical AI workflows in domains with rapidly evolving evidence.

\bibliography{colm2026_conference}
\bibliographystyle{colm2026_conference}

\appendix
\section{Appendix}

\subsection{Corpus construction}

The consensus corpus comprised the following eight expert-curated Long COVID guidance, consensus, and evidence-synthesis documents:
\cite{al2024long,bateman2021myalgic,cheng2025multidisciplinary,mueller2025long,national2024long,peluso2024mechanisms,vogel2024designing,zeraatkar2024interventions}.

We also constructed a ClinicalTrials.gov registry of studies for which people with Long COVID could plausibly enroll or whose results could directly inform Long COVID care. Candidate records were identified through condition-field searches for Long COVID/PASC synonyms and eligibility-text searches for related post-infectious phenotypes, including POTS/dysautonomia, ME/CFS, mast-cell activation, and post-viral cognitive symptoms. Records were retained when they met both a population criterion (Long COVID, prior COVID-19, or a relevant post-infectious phenotype) and a clinical-relevance criterion. Core Long COVID records were retained automatically; other candidates underwent LLM-assisted screening, with uncertain cases reviewed by a clinician. The June 2026 corpus snapshot contained 887 records.

The Living SR corpus was constructed from a monthly PubMed/MEDLINE search combining Long COVID/PASC, randomized-trial, and treatment terms. Eligible records were peer-reviewed randomized controlled trial publications of therapeutic interventions in adults with post-COVID-19 condition. We excluded acute-COVID or prevention trials, anosmia-only studies, nonrandomized or protocol-only reports, and studies with fewer than 25 participants per arm. Candidate records underwent AI-assisted abstract and full-text screening with clinician adjudication. The final corpus contains 74 RCT publications, each represented by structured intervention--outcome profiles linked to GRADE certainty assessments across eight intervention categories.

\subsection{Query-time retrieval}

Table~\ref{tab:retrieval_settings} summarizes the default integrated configuration. Only the consensus corpus uses dense vector retrieval; the remaining sources use query-conditioned structured retrieval.

\begin{table}[t]
\centering
\small
\caption{Source-specific query-time retrieval settings.}
\label{tab:retrieval_settings}
\begin{tabular}{p{0.14\linewidth}p{0.14\linewidth}p{0.31\linewidth}p{0.25\linewidth}}
\hline
\textbf{Source} & \textbf{Corpus} & \textbf{Query-time retrieval} & \textbf{Evidence passed to synthesis} \\
\hline
Consensus&
8 documents; 395 chunks &
FAISS L2-distance search over \texttt{text-embedding-3-large} embeddings &
Top 25 chunks \\
\hline
PubMed &
Live API &
LLM-assisted Boolean query construction, publication filtering, and relevance/quality assessment &
Up to 10 articles \\
\hline
Clinical trials &
887 structured registry records &
LLM-assisted query expansion followed by entity--field ranking over trial metadata; no vector search &
Top 15 records \\
\hline
Living SR &
74 structured RCT profiles&
LLM-assisted query expansion followed by entity--GRADE matching; no vector search &
Top 6, 10, or 14 papers, based on query breadth \\
\hline
\end{tabular}
\end{table}

Selected retrieval arms run independently and are combined in a single synthesis call.

\end{document}